# Binary PSOGSA for Load Balancing Task Scheduling in Cloud Environment

Thanaa S. Alnusairi
College of Computer Sciences and Information
Aljouf University, Skaka, Aljouf

Ashraf A. Shahin[1, 2], Yassine Daadaa[1]
[1]College of Computer and information Sciences
[1]Al Imam Mohammad Ibn Saud Islamic University (IMSIU), Riyadh, Saudi Arabia
[2]Department of Computer and Information Sciences, Institute of Statistical Studies & Research,
Cairo University, Cairo, Egypt

*Abstract*—In cloud environments, load balancing task scheduling is an important issue that directly affects resource utilization. Unquestionably, load balancing scheduling is a serious aspect that must be considered in the cloud research field due to the significant impact on both the back end and front end. Whenever an effective load balance has been achieved in the cloud then good resource utilization will also be achieved. An effective load balance means distributing the submitted workload over cloud VMs in a balanced way, leading to high resource utilization and high user satisfaction. In this paper, we propose a load balancing algorithm, Binary Load Balancing – Hybrid Particle Swarm Optimization and Gravitational Search Algorithm (Bin-LB-PSOGSA), which is a bio-inspired load balancing scheduling algorithm that efficiently enables the scheduling process to improve load balance level on VMs. The proposed algorithm finds the best Task-to-Virtual machine mapping that is influenced by the length of submitted workload and VM processing speed. Results show that the proposed Bin-LB-PSOGSA achieves better VM load average than the pure Bin-LB-PSO and other benchmark algorithms in terms of load balance level.

## I. INTRODUCTION

In the last few years, cloud computing has emerged as a new computing paradigm that primarily aims to provide reliable, customized, and Quality of Service guaranteed dynamic computing environments for end users. Simply, cloud computing is the technology that provides a shared pool of computing resources in the base of on-demand services. In other words, cloud computing is the delivery of computing services such as hosts, storage, databases, networking, software, and more over the Internet. In fact, there are three basic models of services, namely Infrastructure as a Service (IaaS), Platform as a Service (PaaS), and Software as a Service (SaaS). First, in the service model IaaS, a cloud provider delivers datacenters, hosts and virtual machines, storage, networks, and operating systems to cloud users on a pay-as-you-go basis. Second, the service model PaaS delivers services that supply an on-demand environment for cloud users such as developing, testing, delivering, and managing software applications. It is mainly used by application and software developers. Third and finally, the service model SaaS delivers software applications over the Internet to cloud users on-demand and typically on a subscription basis. It is essential to cloud providers to tend the management operations in both task-level and resource-level services. The task-level scheduling allocates a task to a virtual machine (which we address in this study), while the resource-level scheduling allocates a virtual machine to a host. The other important issue is to keep cloud resources balanced. Therefore, they also tend to schedule the incoming application requests to virtual machines in order to complete submitted tasks at the expected time in a balanced way. Numerous objectives have been addressed in the literature, such as minimizing makespan, maximizing load balancing, minimizing flowtime, and minimizing monetary cost.

Considering that task scheduling is NP-complete, many heuristics have already entered the scene, and some have emerged. For instance, Greedy heuristic, Genetic heuristic, Swarm Intelligence-based heuristics such as Ant colony inspired algorithms, Bee Colony inspired algorithms, Fish-inspired algorithms, the Gravitational Search algorithm, and Particle Swarm algorithms. Swarm Intelligence (SI)-based algorithms are population-based and stochastic search algorithms, as these are evolutionary algorithms. In this work, Swarm Intelligence based algorithms are used due to their amazing results achieved in different problems. The SI concept refers to the collective behavior that emerges from the swarms of social insects. Swarms can solve complex problems that exceed the capabilities of their insects without central supervision. What is important, in SI-based scheduling algorithms, is that social insects collectively solve complex problems that are beyond their individual capabilities in an intelligent and decentralized way. As a result, these collective, intelligent, and decentralized behaviors of insects have become a model for solving the problem of task-level scheduling [1], [2]. Due to the impressive performance of SI-based algorithms, researchers have been attracted to this strategy over the last years. Therefore, we have been inspired by the hybrid of a gravitational search algorithm and particle swarm optimization to propose a bio-inspired task scheduling algorithm that can solve the problem of load balancing task scheduling.

By definition, cloud environments are continuously changed since cloud resources are usually dynamically reallocated per demand. This behavior must be captured by the proposed load balancing task scheduling algorithm to manage the process of allocating virtual machines to tasks. The task-level load balancing can explicitly improve makespan, throughput, and scheduling time. On the other hand, resource-level load balancing has also succeeded in increasing the performance in terms of response time, VM migration number and time, and resource utilization [3]. In this paper, we propose a load balancing task scheduling algorithm that has been inspired by the hybrid of particle swarm optimization and gravitational search algorithm to find a near-optimal Task-to-Virtual machine mapping to achieve best-balanced resource allocation as well as minimize makespan. The proposed algorithm is a dynamic load balancing algorithm that is more suitable for cloud environments due to its dynamic nature. Although the dynamic load balancing algorithm can consider all changes during runtime, it achieves better results than static algorithms [4].

The rest of the paper is organized as follows. Section II discusses related works. Sections III, IV, and V present a brief introduction to the standard GSA, Standard PSO, and Standard PSOGSA algorithms, respectively. Section VI explains the proposed binary load balancing PSOGSA. Section VII discusses the complexity of the proposed algorithm. Section VIII describes the neighborhood topology of the proposed algorithm. The objective function is discussed in Section IX. Section X explains in detail the technical processes of the proposed algorithm. The experimental results are presented in Section XI. Finally, Section XII concludes the work and suggests some directions for future work.

## II. RELATED WORK

This literature review presents many works under the umbrella of Load Balancing to resolve problems related to different performance parameters, such as throughput, CPU utilization, overhead, fault tolerance, migration time, number of migrations, response time, and makespan. Some of these parameters involve the efficiency of task-level schedulers, and some involve resource-level schedulers.

ParticleZ in [4] solves the problem of task scheduling in grids through four phases: job submission, queuing, node communication, and job exchanging. In particular, the role of PSO appears in the phases of communication and exchange. In those phases, particles (nodes) search to find the best position (node with minimum load among neighbor positions). Further, to guarantee load balancing all the time, each node (particle) exchanges its loads with its neighbors in a parallel way. In load search space, the lower the load is, the higher the velocity of the particle. On the other hand, to establish a fair load distribution between each neighbor's set, the exchanged loads have to be under a predefined threshold (second lightest load of neighbor's set - lightest load).

Aslanzedh et al. [5] have been inspired by the endocrine system to improve the load balancing technique in cloud management. In fact, they have combined the endocrine system and PSO (Endocrine-PSO), aiming to schedule tasks as well as minimize makespan in load-balanced resources. The Endocrine-PSO algorithm has employed the functions of hormone regulation (load balance side) and PSO (task scheduling side) to perform the scheduling process efficiently. By endocrinology science, there is the push-pull procedure, which describes the hormone regulation process in the human body (push means stimulating a hormone from a gland, while pull means inhabitation the hormone to another gland). Technically, Endocrine-PSO provides two particles: one for the push operation (St), and the other for pull (Dt). The St and Dt carry values of stimulating hormones and inhibiting hormones respectively. Results show that the Endocrine-PSO can find the best mapping, either for choosing the best task-to-VM schedule or migrating the tasks from overloaded under-loaded VM.

Different criteria have been evaluated in [6] for the PSO-based scheduling algorithm that has developed to increase the efficiency of the load balancing scheduling process in clouds. This algorithm (LBMPSO) aims to minimize makespan, transmission time, and transmission cost, and to maximize reliability and load balancing. LBMPSO guarantees the reliability of clouds by rescheduling tasks that have failed to be scheduled as well as guarantees load balancing between tasks and available VMs.

Jena [7] has proposed a Nested and Multi-objective PSO Framework for task scheduling in a cloud environment. Furthermore, other criteria have been addressed in this paper. The MOPSO (Multi objective PSO) algorithm has been proposed to minimize energy and makespan. The author has hybridized PSO with an evolutionary algorithm to create the proposed multi-objective algorithm. Additionally, another concept has been considered to make solutions of MOPSO valuable spread solutions that are selected based on Pareto dominance.

Dasgupta et al. [8] have modified the load balancing genetic algorithm with a new objective function that guarantees the user's QoS preferences as well as minimizes response time. The authors have contributed the weights of the objective function to satisfy users' preferences. This algorithm outperforms its rivals such as SCH, RR, and FCFS. However, a limitation in this load balancing algorithm has been observed: it considers that all jobs have the same priority, which is not the case of real-world jobs.

Xin Lu and Zilong Gu [9] have proposed an ACO-inspired load-adaptive cloud resource scheduling algorithm to maximize CPU utilization. It has solved two issues, the detection of hotspot node (the overloaded VM) and adaptive resource scheduling. The proposed model would monitor the CPU usage, memory, and bandwidth of all VMs within a cluster, and if a hotspot VM is detected, the scheduling process starts. The resource scheduling process is performed to find the idle node that contributes to lighten the load over the hotspot node. The authors added an expansion factor to the global update to enable faster convergence of the ant to the path that has the desired resources by expanding its pheromone intensity. Results show that the proposed algorithm easily detects overloaded VMs and finds the nearest idle node.

More related research has addressed the load balancing issue in grid environments. Ludwig et al. [4] have introduced the AntZ approach. They have enhanced the previous works [10] and [11] to adopt the problem of load balancing efficiently. They take advantage of decay rate in [11] and mutation rate in [10] and then combine them into AntZ. Specifically, the mutation rate addresses the problem of load balancing due to its effect on guiding ants to the best path.

Dhinesh Babu and Venkata Krishna [12] have proposed a honey bee inspired load balancing algorithm (HBB-LB). The proposed HBB-LB strategy schedules tasks by taking into consideration the VMs' load balance aiming to minimize makespan, response time, and number of migrations. The proposed algorithm divides VMs based on its computing capacity into overloaded VMs, under-loaded VMs, and balanced VMs. The balancing process is performed by removing tasks from the overloaded VMs and submitting them to the under-loaded VMs with respect to task priority. The removed tasks act as honeybees, and the under-loaded VMs represent a profitable nectar source. Results show that the HBB-LB algorithm works robustly without heavy overhead, and also works efficiently in heterogeneous cloud systems.

Lili Xu and Kun Wang in [13] have proposed a green cloud task scheduling algorithm (GCTA) based on an improved binary PSO variant. In the proposed algorithm, they tried to enhance the binary PSO solution by avoiding matrix operations and using pipelined numbers for virtual machines. The authors have compared the performance of the proposed GCTA algorithm with a sequential scheduling algorithm and found that the proposed GCTA strategy achieves better performance.

## III. STANDARD GSA

The GSA treats masses as search agents. The Newtonian laws of gravity and motion define how all masses move in the direction of other masses and the speed at which they do so. The greater the mass, the slower the movement and the greater the attraction to the other masses is. Since, in the GSA, a greater mass means a better solution, the GSA is seen as an excellent way to guarantee convergence with the optimum. Every mass has a position; it also has inertial, active gravitational and passive gravitational masses. Theoretical physics defines these properties in the following way [14]:

Active gravitational mass: measures how strong an object's gravitational field is. Objects with small active gravitational mass have weaker gravitational fields than objects with greater active gravitational mass.

- Passive gravitational mass: measures how strong is the interaction of an object with the gravitational field. Objects with small passive gravitational mass are subject to a weaker force than objects in the same gravitational field with larger passive gravitational mass.

- Inertial mass: measures the strength of the resistance offered by an object to changes in its motion state as a result of the application of force. Objects with large inertial mass will undergo a slower state change as a result of the application of force than objects with small inertial mass.

If we assume the existence of s masses, then the position vector of the k$^{th}$ mass object at time(t) $X_k(t)$ will be as set out in (1):

$$X_k(t)=\{x_1^k \quad x_2^k \quad \cdots \quad x_i^k \quad \cdots \quad x_n^k\} \quad (1)$$

As well as positional property, each mass also possesses velocity and acceleration, and these may be represented using a vector.

The acceleration vector $Acc_k(t)$ of mass object k at time t is a vector of n elements as follows:

$$Acc_k(t)=\{acc_1^k \quad acc_2^k \quad \cdots \quad acc_i^k \quad \cdots \quad acc_n^k\} \quad (2)$$

while the velocity vector also has n elements as follows:

$$V_k(t)=\{v_1^k \quad v_2^k \quad \cdots \quad v_i^k \quad \cdots \quad v_n^k\} \quad (3)$$

Additionally, the vector of global best positions at time t is:

$$Xgbes_t(t)=\{xgbest_1^k \quad xgbest_2^k \quad \cdots \quad xgbest_i^k \quad \cdots \quad xgbest_n^k\} \quad (4)$$

Equation (5) sets out the force exerted on object i by the object j:

$$F_d^{ij}(t) = G(t) + \frac{M_i(t) \times M_j(t)}{R_{ij}(t) + \epsilon} (x_{id} - x_{jd}) \quad (5)$$

In this equation, $M_j$ is the value of the mass related to mass object j, $M_i$ is the value of the mass related to mass object i, ε is a small constant, and $R_{ij}(t)$ is the straight-line distance in Euclidean space between mass object i and mass object j. Equation (6) gives us the value of G(t) as a function of initial value G0 at iteration t:

$$G(t) = G_0 \times exp\left(\frac{-\alpha \times t}{t_{max}}\right) \quad (6)$$

In this equation, G0 is the initial gravitational constant and α is a user-defined descending constant, t is the current iteration, and $t_{max}$ is the maximum number of possible iterations. $F^i_d(t)$ is the total force exerted in the d$^{th}$ direction on mass object i and is a sum (randomly weighted) of the d components of other mass objects' forces:

$$F_d^i(t) = \sum_{j=0}^{n} rand_i \times F_d^{ij}(t) \quad (7)$$

In this equation, $rand_i$ is a uniform random variable in the interval [0, 1].

The object i accelerates in the d$^{th}$ direction at time t at the rate of $Acc^i_d(t)$, calculable according to (8):

$$Acc_d^i(t) = \frac{F_{i,d}(t)}{M_{ii}(t)} \quad (8)$$

where $M_{ii}$ is the mass object i inertial mass. Equations (9) and (10) calculate, respectively, this object's next velocity and position at time t+1:

$$v_d^i(t+1) = rand_i \times v_d^i(t) + acc_d^i(t) \quad (9)$$
$$x_d^i(t+1) = x_d^i(t) + v_d^i(t+1) \quad (10)$$

where $rand_i$ is, once again, a uniform random constant in the interval [0, 1]. Its purpose is to give the search a randomized characteristic. Current velocity and current position are, respectively, expressed as $v^i_d(t)$ and $x^i_d(t)$.

## IV. STANDARD PSO

The usual use of the population-based algorithm PSO (Particle swarm optimization) is the efficient solution of problems of optimization, and PSO is one of various techniques of swarm intelligence used to solve problems of optimization.

In this class of techniques, "particles" (search agents) fly in the optimization problem's search space. This activity is a representation of the process of searching – it is, in effect, a journey that searches for the best position that can be taken by a particle. Each search agent, or particle, is a candidate for the role of optimal optimization problem solution, and each changes velocity and position to look for an improved position in the search space. These changes of velocity and position follow the rules deduced originally from behavioral models representing the flocking of birds as proposed by Kennedy and Eberhart in [15]. In each case, calculation of next velocity and position (respectively $v_{i,d}(t+1)$ and $x_{i,d}(t+1)$) is specified (again respectively) by (11) and (12):

$$v^i_d(t+1) = (w \times v^d_i(t)) + (c_1 \times rand_i \times (pbest^i_d(t) - x^i_d(t)))$$
$$+ (c_2 \times rand_i \times (gbest^i_d(t) - x^i_d(t))) \quad (11)$$

$$x^i_d(t+1) = x^i_d(t) + v^i_d(t+1) \quad (12)$$

Once again, $rand_i$ is a uniform random constant in the interval [0,1] and is used to randomize the search. The $pbest^i_d$ represents the current mass object's personal best position on the $d^{th}$ direction, while $gbest_d$ represents the ith mass object's global best position on the $d^{th}$ direction at iteration t. Current velocity and position are represented respectively by $v_{i,d}(t)$ and $x_{i,d}(t)$.

In fact, Kennedy and Eberhart have proposed another variant of PSO. In [16], they have proposed the binary version of PSO that is proposed to solve discrete problems. The significant difference in binary PSO is the way in which positions can be updated. Updating of positions is specified by finding the value of the sigmoid function for each mass' velocities as in the following (13):

$$Sig(v^{t+1}_{i,d}) = \frac{1}{1 + exp(-v^{t+1}_{i,d})} \quad (13)$$

Values that returned from the sigmoid function are normalized, as defined in (14):

$$x^k_{ij} = \begin{cases} 1, & \& rand_i < v^{t+1}_{i,d} \\ 0, & Otherwise \end{cases} \quad (14)$$

where $rand_i$ is a uniform random constant in the interval [0,1]. Here, the sigmoid function is used to transfer a real-valued velocity $v_{i,d}$ to a probability value in the range of [0, 1] [23].

## V. STANDARD HYBRID PSOGSA

The Hybrid PSOGSA metaheuristic is a low-level bio-inspired heterogeneous hybrid algorithm. Seyedali Mirjalili and Siti Zaiton Mohd Hashim have proposed the Hybrid PSOGSA in [17] as a novel algorithm. In fact, they have hybridized the standard PSO, and Standard GSA mentioned in the last two sections, to balance the exploration and exploitation abilities of GSA and PSO. The core idea of the Hybrid PSOGSA is to combine the exploration of GSA and the exploitation of PSO.

In other words, the strong points of both PSO and GSA were taken into consideration to improve the weakness of GSA exploitation ability as well as PSO exploration. As tested in [17], the Hybrid PSOGSA has very good exploration and exploitation abilities, which are due to its ability to avoid becoming stuck in local optima and tending to converge to the best solution quickly.

The combination of PSO exploitation and GSA exploration is translated into a new velocity equation (1). That Hybrid PSOGSA velocity integrates the velocity of both GSA and PSO to boost the balance between global search capability of GSA and local search capability of PSO. The Hybrid PSOGSA velocity equation considers the acceleration of the mass object rather than pbest as in PSO velocity, which indicates that the Hybrid PSOGSA relies on the global search of PSO with the local search of GSA. The velocity of mass object i on the $d^{th}$ dimension at next iteration (t+1) is $v_{i,d}(t+1)$ and its position $x^i_d(t+1)$ is calculated according to (15) and (16), respectively:

$$v^i_d(t+1) = (w \times v^d_i(t)) + (c_1 \times rand_i \times acc^i_d(t))$$
$$+ (c_2 \times rand_i \times (gbest^i_d(t) - x^i_d(t))) \quad (15)$$

$$x^i_d(t+1) = x^i_d(t) + v^i_d(t+1) \quad (16)$$

where $rand_i$ is a uniform random constant in the interval [0,1]. This random number is used to give a randomized characteristic to the search, $acc^i_d(t)$ is the acceleration of the current mass object on the $d^{th}$ direction, and $gbest_d$ is the global best position of $i^{th}$ mass object on $d^{th}$ direction in iteration t. The $v_{i,d}(t)$ and $x_{i,d}(t)$ are its current velocity and position, respectively.

A good balance between exploration and exploitation can be achieved by controlling terms of the velocity equation based on its factors w, c1, and c2. The functions of these terms and these factors are explained as follows:

*1) Momentum component ($w \times v_{id}(t)$):* The inertial factor w characterizes inertia of masses, i.e., it controls the momentum of masses and how much mass remembers its previous velocity. Larger w causes the mass to have a better exploration ability, and smaller w values allow the mass to have a better exploitation ability.

*2) Cognitive component ($c1 \times rand_i \times acc^i_d(t)$):* The first behavioral factor c1 controls how much mass can be influenced by its acceleration at iteration t.

*3) Social component ($c2 \times rand_i \times (gbest - x_i(t))$):* The second behavioral factor c2 controls how much a mass can head toward the population's best mass.

In the case of c1 and c2, the larger values cause the mass to have a better exploitation ability. In fact, effective values can permit these three factors to achieve a good balance between exploration and exploitation.

## VI. Proposed Binary Load Balancing PSOGSA

The Bin-LB-PSOGSA (Binary Load Balancing PSOGSA) works to distribute submitted application requests over VMs in an efficient, balanced distribution. At each time, requests of different users are submitted at different submission times to the cloud system. Then, a search process is performed by Bin-LB-PSOGSA to assign tasks of submitted application requests to VMs in a dynamic way. At the same time, a rescheduling of tasks that have already been submitted is re-applied, i.e., the task is bound to the submitted requests list, and a new search process is performed based on the new submitted requests list.

Unlike continuous search space, the search space is represented as a hypercube. Each mass moves over hypercube nodes (corners) by flipping one or more bits of the mass position matrix. Iteratively, the position matrix of a mass is binary-coded. But, the velocity matrix still consists of continuous values belonging to the real numbers. Each velocity element value holds the probability to flip or change the binary value of the corresponding position element. The process of flipping (or changing binary value) is performed by using a transfer function. In fact, transfer functions are used to determine the probability of the value of each bit in the mass position matrix (0 or 1). [18]

In the deep search process of Bin-LB-PSOGSA, each mass in the population represents one candidate solution or, in other words, a task-to-VM mapping. Each candidate solution has a fitness value, which is the value of the expected finish time of each submitted application request.

In pseudo code of the Bin-LB-PSOGSA (see Algorithm 1), first, the masses' population is initialized by the function initialize Masses (tasks, VMs) at line 2. In the initialization phase, tasks are assigned randomly to VMs. Then, for each iteration, global variables are updated that have to be changed iteratively, such as gravitational constant, best mass, and worst mass, by updateGlobalVariables(iteration) at line 4. Until the maximum iteration is reached and for each mass object in the population, the mass value of each mass object and gravitational force exerted by the population masses is calculated at lines 6 and 7, respectively. Then, at lines 8 to 10, mass position, velocity, and fitness are updated. As in Algorithm 2, the pseudo code that has clarified the way to update the velocity of each mass is presented.

After updating the mass's fitness, it is necessary to decide if the new fitness is better than what the mass object already found in its trip; this is done in line 12. If the new fitness is better, the personal best fitness and mass will be updated. Consequently, if the personal best mass is better than all solutions found by all mass objects, the global best mass is updated as in line 16. In fact, the global best mass is the promising best mass object that attracts most of the population due to its mass value (biggest mass value or heaviest mass). Based on the topology of the neighborhood that has been considered in this variant, Gbest topology (discussed in subsection VIII), the population is influenced by the best global mass positions that are updated at line 18. In the next iteration, by updating the best global positions so far, other masses' objects take their new positions. Over time, most of the population comes increasingly closer to the best global mass, and finally, if the maximum iteration number is reached, the search process is terminated. Eventually, the best global mass is returned in the form of the best task-to-VM mapping at line 22.

## VII. Bin-LB-PSOGSA Complexity

Let s be population size, v be VMs size, and c be submitted requests' tasks size. Initialization of masses is used to add random positions and velocities of each mass in the population. During initialization, the fitness of the current position of the mass is calculated. The time complexity of mass initialization is $O(v \times c)$. So, the time complexity for initialization of the whole population is $O(s \times v \times c)$.

In the iterations loop, first, global variables are being updated. The time complexity of that action is $O(s^3 + s \times c)$. The reason for such time complexity is that inside update Global Variables (iteration) there is a need for collecting the best and worst fitnesses from the whole swarm. The time complexity of those actions is $O(s)$. Inside this method, we also calculate the total forces that act on each mass and acceleration of the mass. The time complexity of those actions is $O(s^3 + s \times c)$. This is also the time complexity of the method update Global Variables (iteration).

Second, a loop is iterated for each mass in the population. Each mass's velocity and position are being updated. The time complexity of both of these updates is $O(v \times c)$. The next step is to update the fitness of the mass. The time complexity of that action is $O(v)$. The overall time complexity of the particle loop is $O(s \times v \times c)$.

For the iterations loop, the time complexity is therefore $O(s \times MAX\_ITERATION \times (s^2 + v \times c))$.

The final step is to return the mapping from cloudlet to VM. The time complexity of this action is $O(v \times c)$. Eventually, if the time complexity of each step is combined, the final result is $O(s \times MAX\_ITERATION \times (s^2 + v \times c))$.

## VIII. Neighborhood Topology

The neighborhood topology adopted in the proposed Bin-LB-PSOGSA is the global neighborhood topology (Gbest) [19]. In other words, Gbest is a fully connected topology where all the masses are neighbors of each other and able to exchange information with each other. Further, the process of exchanging is fast due to the full connection between all population masses. Gbest topology makes the proposed Bin-LB-PSOGSA a fully informed strategy where every mass in the population learns from the same global best mass and is influenced by its positions.

```
Algorithm 1: Bin-LB-GSAPSO
1  function balance(tasks, vms)
     Input: tasks - list of submitted requests' tasks; vms - list of VMs
     Output: task to VM mappings
2    masses ← initializeMasses(tasks, vms);
3    for iteration ← 1 to MAX_ITERATION do
4      updateGlobalVariables(iteration);
5      foreach mass in masses do
6        calculateMass(mass, population);
7        calculateRelevantForce(mass, population);
8        updateVelocities(mass);
9        updatePositions(mass);
10       fitness ← updateFitness(mass);
11       personalBestFitness ← mass.getPersonalBestFitness();
12       if fitness < personalBestFitness then
13         personalBestFitness ← fitness;
14         personalBestPositions ← mass.getPositions();
15       end
16       if personalBestFitness < globalBestFitness then
17         globalBestFitness ← personalBestFitness;
18         globalBestPositions ← personalBestPositions;
19       end
20     end
21   end
22   return taskToVmMapping(globalBestPositions);
23 end
```

```
Algorithm 2: Update velocity
1  function updateVelocities(mass)
     Input: mass - current mass
2    for vm to vms do
3      vmVelocities[ ] ← mass.getVelocities()[vm.id];
4      vmPositions[ ] ← mass.getPositions()[vm.id];
5      for task to submittedTasks do
6        newVelocity[vm.id][task.id] = (w * vmVelocities[task.id])
7        + (c_1 * Random.nextDouble() * currMass.getAcceleration()[task.id])
8        + (c_2 * Random.nextDouble() * (vmGlobalbestPositions[task.id]
           - vmPositions[task.id]));
9      end
10   end
11 end
```

## IX. OBJECTIVE FUNCTION

The solution of proposed algorithms is to minimize the expected execution time of each task in submitted application requests or $ET_{ij}$ of the task $T_i$ that is running on $VM_j$. The calculation of $ET_{ij}$ is as follows:

$$ET_{ij} = \frac{TL_i}{PS_{ij}} \quad (17)$$

The processing speed $PS_{ij}$ related to $T_i$ running on $VM_j$ in the cloud depends on how many request tasks have been mapped to that VM (or n) as well as the total allocated MIPS of $VM_j$ along all its processing elements or Pes (or $Capacity_j$). Calculation of the request processing speed $PS_i$ is shown in (18):

$$PS_{ij} = \frac{Capacity_j}{n} \quad (18)$$

## X. BIN-LB-PSOGSA TECHNICAL PROCESSES

In this section, the mathematical notations and technical processes' steps are discussed. In the proposed algorithms, binary search space is considered. Therefore, the binary matrix encoding form in [13] is adopted to represent mass objects. Accordingly, each mass object has properties of mass position matrix and mass velocity matrix that will be decoded to a two-dimensional matrix. The first dimension represents the VM number and the other tasks number. The position matrix takes binary values, while the velocity matrix keeps the continuous values. Additionally, each corresponds to a task-to-VM mapping as a candidate solution.

Here, the mathematical notation of the problem is described. The task set T = { T1,T2, ….. ,Ti, …… , Tc } will be mapped to VMs set VM = { VM1,VM2, ….. ,VMj, …… , VMv } using the relationship matrix representation in (18). Let s be the masses' population size, each mass object m represented by position matrix Xm of c × v position elements ( c is the number of tasks and v is the number of VMs where i = {1, 2, 3, …. , c} and j = {1, 2, 3, ….. , v} and m = {1 , 2 , 3 , ….. , s}) as given in follows:

$$X_m = \begin{pmatrix} x_{11}^m & x_{12}^m & \cdots & x_{1c}^m \\ x_{21}^m & x_{22}^m & \cdots & x_{2c}^m \\ x_{v1}^m & x_{v2}^m & \cdots & x_{vc}^m \end{pmatrix}$$

The element $x^m_{ii}$ is the position of mass object *m* in row *i* and column *j*; actually, it represents the distribution relationship between task $T_i$ and virtual machine $VM_i$ i.e., it explains whether $T_i$ is mapped to $VM_i$ or not. Position $x^m_{ii}$ takes values of either 0 or 1. Namely, it indicates on which VM task $T_i$ is working. So, if $T_i$ is running on $VM_i$ then position $x^m_{ii}$ is equal to 1, but it equals 0 otherwise. Finally, positions $x^m_{ii}$ that are equal to 1 are recorded composing solution position vector $P_m$ (t) at time *t* as in (19):

$$P_m(t) = \{p_{1,j}^m \quad p_{2,j}^m \quad \cdots \quad p_{i,j}^m \quad \cdots \quad p_{c,v}^s\} \quad (19)$$

where $p^m_{i,j}$ can take one of $x^m_{ii}$ of the relation distribution matrix that has a value equal to 1 and j = (1, 2, … , v).

Here, the technical steps of the proposed algorithms are explained in detail.

### A. Population Initialization

Initially, position elements $x^m_{ij}$ of each mass m position matrix $X_m$ are initiated randomly by mapping each task to random VM, i. e. for each column at index i, one element is arbitrary assigned to the value 1 and the remaining elements (in that column) to 0. Iteratively, this process – initiating mass position matrix -- is repeated c×v times for each mass in the population.

For instance, assume that there are seven tasks and three VMs, and the population consists of 50 masses; therefore, the initial position matrix of mass 20 ($X_{20}$) will be as follows:

$$X_{20} = \begin{pmatrix} 0 & 1 & 0 & 0 & 0 & 1 & 0 \\ 1 & 0 & 1 & 1 & 0 & 0 & 1 \\ 0 & 0 & 0 & 0 & 1 & 0 & 0 \end{pmatrix}$$

Namely, it is shown in the matrix ($X_{20}$) that $T_1$ is mapped to $VM_2$, $T_2$ is mapped to $VM_3$, $T_3$ is mapped to $VM_2$, $T_4$ is mapped to $VM1_2$, $T_5$ is mapped to $VM_1$, $T_6$ is mapped to $VM_1$, and $T_7$ is mapped to $VM_3$.

Finally, the tasks-to-VM vector or dimension vector $D_k$ is defined in which indices of tasks' positions that hold value 1 are stored in it respectively as in (20):

$$D_k = \{d_1^k \quad d_2^k \quad \cdots \quad d_i^k \quad \cdots \quad d_c^k\} \quad (20)$$

where $d_t^k$ is the index of assigned task $T_i$ in position matrix $X_k$ of the mass k, and vector indices are the VMs' IDs.

### B. Mass Value, Relevance Force, and Acceleration Calculation

In this step, a mass value, relevance force, and finally the acceleration are calculated based on the gravity law equations in [14]. First, the mass value $M_m(t)$ of each mass object at iteration t is calculated based on (21) and (22):

$$mass_m(t) = \frac{fitness_m(t) - worst(t)}{best(t) - worst(t)} \quad (21)$$

$$M_m(t) = \frac{mass_m(t)}{\sum_1^{s-1} mass_b(t)} \quad (22)$$

where $fitness_m(t)$ is the fitness value of the mass m at iteration t, $worst(t)$ is the global worst fitness in iteration t, and $best(t)$ is the best one at that iteration. Further, $mass_m(t)$ is the current mass value where $mass_b^m(t)$ is a vector that holds the mass values of neighbor masses' objects at iteration t as shown in (23). This is because $\sum_1^{s-1} mass_b(t)$ is the summation of other masses' values. In fact, global best fitness represents the minimum expected finish time that can be achieved by the best global task-to-VM mapping since the global worst is the longest expected finish time.

$$mass_b^m(t) = \{mass_{b,1} \quad mass_{b,2} \quad \cdots \quad mass_{b,(s-1)}\} \quad (23)$$

Second, the relevance gravitational forces exerted on mass m by another mass b is calculated based on (24):

$$F_d^{m,b}(t) = G(t) + \frac{M_m(t) \times M_b(t)}{R_{mb}(t) + \epsilon}(x_d^m - x_d^b) \quad (24)$$

In this equation, $M_b$ is the value of the mass related to the active gravitational mass of object b, $M_m$ is the value of the mass related to passive gravitational mass of object m, ε is a small constant, and $R_{mb}(t)$ is the straight-line distance in Euclidean space between mass object m and mass object b, and $x_d^m$ and $x_d^b$ are the corresponding positions at dimension d in both of the passive masses m and b. It is useful to mention that the active mass is the mass that generates the gravity, while the passive mass the mass responds to the gravity. The relevant gravitational force values are defined in a vector as in (25):

$$F_d^{m,b}(t) = \{f_1^{m,b} \quad f_2^{m,b} \quad \cdots \quad f_i^{m,b} \quad \cdots \quad f_c^{m,b}\} \quad (25)$$

where $f_i^{m,b}$ is the value of the exerted force on passive mass m from active mass b in the dimension d (number of dimensions d equals the number of tasks c).

Then, the total result gravitational forces exerted on mass m on the $d^{th}$ direction at time (t) are as in (26):

$$F_d^m(t) = \sum_{k=1, k \neq m}^{s-1} rand_k \times F_d^{m,b}(t) \quad (26)$$

In this equation, $rand_m$ is a uniform random variable in the interval [0, 1] generated for each mass m.

Based on Newton's law of gravity and Newton's law of motion, each mass object m moves toward the global best mass object by updating the acceleration vector of that object iteratively. Under the concept of motion law, the acceleration of the mass object m on the $d^{th}$ direction at time t is $Acc_{m,d}(t)$ as in the following equation:

$$Acc_d^m(t) = \frac{F_{m,d}(t)}{M_{mm}(t)} \quad (27)$$

where $M_{mm}$ is the mass object m inertial mass.

### C. Velocity Updating

The binary version of the velocity equation is as follows:

$$V_m(t+1) = (w \times v_{ij}^m(t)) + (c_1 \times rand_m \times acc_i^m(t))$$
$$+ (c_2 \times rand_m \times (gbest_{ij}^m(t) - x_{ij}^m(t))) \quad (28)$$

where $V_m(t + 1)$ is the velocity matrix of mass object m for the next iteration, and $v_{ij}^m(t)$ is the current velocity value of the element related to $T_i$ and $VM_i$. The acceleration of mass m is $acc_i^m(t)$. The inertial weight (w) is calculated based on (29) where acceleration coefficients C1 and C2 are based on, respectively, (30) and (31). The $rand_m$ is a uniform random constant in the interval [0, 1] generated for each mass object m. Its purpose is to give the search process a randomised characteristic.

For each iteration (t), mass object m records the best global positions in its memory so that all masses can be increasingly closer until a maximum number of iterations is reached. Iteratively, the distance between masses and global best mass is decreased by subtracting the distance between positions $x^m_{ij}(t)$ and $gbest^m_{ii}(t)$, as stated in equation 28, term $(gbest^m_{ii}(t) - x^m_{ii}(t))$. Elements $Xgbest_m$ and current mass position matrix are subtracted one by one. In the case of the acceleration of masses' objects, a constant random value $(c_1 \times rand_m)$ is multiplied with all elements in the acceleration vector, element by element. Also, in the current velocity matrix, the current value of the inertia weight is multiplied with all elements in the velocity matrix. The velocity matrix will be as follows:

$$V_m = \begin{pmatrix} v_{11}^m & v_{12}^m & \cdots & v_{1c}^m \\ v_{21}^m & v_{22}^m & \cdots & v_{2c}^m \\ v_{v1}^m & v_{v2}^m & \cdots & v_{vc}^m \end{pmatrix}$$

To enhance the search process, we have considered a time-adaptive approach for the other controlling parameters such as inertia weight and acceleration coefficients ($c_1$ and $c_2$) as in (28), (29), and (30). For the inertia weight, we have adopted a time-varying inertia weight as introduced in [20] and acceleration coefficients in [21]. Here, $w_{max}$ and $w_{min}$ have constant values equal to 0.9 and 0.4, respectively, t is the current iteration, and $t_{max}$ is the maximum iteration.

$$w = \frac{w_{max} - w_{min}}{t} \quad (29)$$

$$c_1 = 1 - \frac{t^3}{t_{max}} \quad (30)$$

$$c_2 = \frac{t^3}{t_{max}} \quad (31)$$

### D. Position Updating

Each mass moves to the global best mass by updating positions and becomes increasingly closer to the global best mass over iterations.

Also, for the transfer function in the proposed Bin-LB-PSOGSA, we have used a time-adaptive approach as introduced in [22]. The time-varying transfer function is used here to enhance the exploration and exploitation processes. Moreover, to transform a real-valued velocity $V_M$ to a binary value (0 or 1) in the process of updating positions (re-encoding) values of relation distribution matrix elements $x^k_{i,i}$ [13]. In fact, if the absolute value is large, the probability to flip a bit is higher. Updating of the position matrix elements is performed by applying the time-varying transfer function ($TV_t$) for each $v^m_{ij}$ in mass' velocity matrix as stated in (32) and (33).

$$TV_t(t) = \frac{1}{1+e^{\frac{v^m_{ij}}{\varphi}}} \quad (32)$$

$$\varphi = \varphi_{max} - t \times \frac{(\varphi_{max} - \varphi_{min})}{t_{max}} \quad (33)$$

where $\varphi_{max}$ and $\varphi_{min}$ have constant values equal to 1.0 and 5.0, respectively.

*E. Finding Global best Task-to-VM mapping*

By the end of each iteration t, the best task-to-VM maps have been read from the global best mass position matrix. This process happens t times and for only the global best mass found during the searching process.

## XI. EVALUATION

In this section, we show how to evaluate the proposed Bin-LB-PSOGSA to test its efficiency in achieving cloud balancing in term of the submitted load. The next subsections discuss the simulation tool and simulation setup we have used in the experiments. Additionally, we explain the algorithm meta-parameters, and finally, we conclude with the results of these experiments. In the last subsection, we assess the performance of the proposed algorithm in terms of load average and processing speed average against the Bin-LB-PSO algorithm.

*A. Experimental Tool*

The performance analysis of the proposed algorithm is carried out in a cloud simulator. The simulator CloudSim [23] is one of the best simulators for experimental purposes. This simulator is a generalized simulation framework that allows modeling, simulation, and experimenting with cloud computing infrastructure and application services.

In this section, we have analyzed the performance of our algorithm based on the results of simulation done using CloudSim. We have extended the classes of the CloudSim simulator to simulate our algorithm.

*B. Simulation Setup*

The simulation setup is detailed in Tables I and II. The experiment is carried out with 3 Datacenters each having two hosts, and the characteristics are 1024 MIPS Host processing power, 2 GB RAM, 1000 GB storage, 10240 Mbps (bandwidth), and 2 PEs (or cores). Each PE had the same processing power, as clarified in Table I.

In Table II, there are 5 VMs, and the characteristics are 128 MIPS (VM processing power) and 2 PEs (or cores).

In this experiment, the workload has been selected as introduced in [24]. Each task in the workload log, called a cloudlet by Cloudsim, was determined by the parameter PEs, or the number of processing elements (cores) required to perform each task. Each cloudlet required 4 to 256 PEs. The number of PEs is limited to powers of 2 due to the architecture of the supercomputer used in the log.

TABLE I. DATACENTER CONFIGURATION

| | |
|---|---|
| Number of datacenters | 1 |
| Number of hosts per datacenter | 4 |
| Number of PEs per host | 1 |
| Number of MIPS per PE | 1024 MI |
| RAM | 2048 MB |
| Storage | 1048576 MB |
| Bandwidth | 10240 MB/s |

TABLE II. VM CONFIGURATION

| | |
|---|---|
| Number of VMs per host | 5 |
| Number of PEs per VM | 2 |
| Number of MIPS per PE | 128 |

*C. Algorithm Meta-Parameters*

The algorithm meta-parameters, or in other words, the controlling parameter settings of Bin-LB-PSOGSA, are as mentioned in equations 28, 29, and 30. The maximum number of iterations is 500, and population size (number of masses) is 50. The acceleration constants C1 and C2 are set to 2 and 2, the inertial weight is linearly decreasing from 0.9 to 0.4. The initial gravitational constant (G0) is 1, descending constant (α) is 20, and small gravitational constant (ε) is $e^{-1}$. The search space bounds are in the range [0, 100], and the velocity range is [-8, 8].

*D. Experimental Results*

Here, the performance of the proposed algorithm in terms of average VM load and average VM processing speed is discussed. The next subsections explain the performance from different sides in detail.

*1) Average VM load over time*

Fig. 1 shows that in comparison with Bin-LB-PSO, the load of the proposed Bin-LB-PSOGSA is smaller than the load of Bin-LB-PSO in general. In particular, both Bin-LB-PSO and the proposed Bin-LB- PSOGSA have stable load values for a long time. This situation is due to the stability of the system because of the number of running application requests. After time passes, both as the time passes, the load increases due to the growth in the number of running application requests. Under the same environmental conditions, the proposed Bin-LB-PSOGSA outperformed Bin-LB-PSO in keeping the system balanced much longer. For instance, at moment 28, the load value of Bin-LB-PSO leaps due to the gap of some successes of application requests before and after this moment (before it was 2 and after is 61). On the other hand, the load average in the proposed Bin-LB-PSOGSA has lower load average values than Bin-LB-PSO.

*2) Average VM processing speed over time*

Fig. 2 shows that over time, the average processing speed of application requests of both algorithms decreases in general. In particular, at moment 28, processing speed decreased dramatically due to the obvious increase in the running requests, which indicates the efficiency of both algorithms to utilize VMs. Therefore, it is clear that both Bin-LB-PSO and the proposed Bin-LB-PSOGSA have utilized VMs efficiently over time.

Both Fig. 1 and 2 shows that as the load increases over time, the processing speed of the submitted applications decreases, which proves that the proposed Bin-LB- PSOGSA is more efficient in keeping the load balanced over time as shown in Table III.

TABLE III. PERFORMANCE COMPARISON

| Criteria | Bin-LB-PSOGSA | Bin-LB-PSO |
|---|---|---|
| VM processing speed (MIPS) | 4376.84346 | 4376.84346 |
| VM Load | 389.0400819 | 451.3841267 |
| Expected execution time (ms) | 3900000 | 3800000 |
| Performance efficiency | Better | Limited |

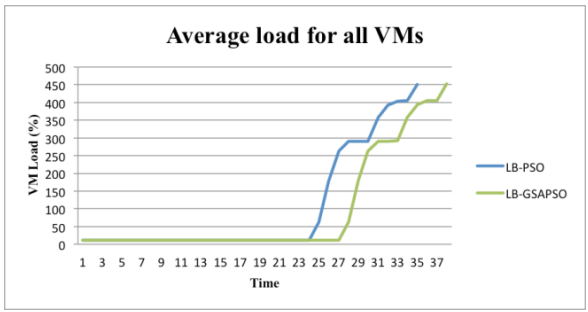

Fig. 1. Average VM load over time.

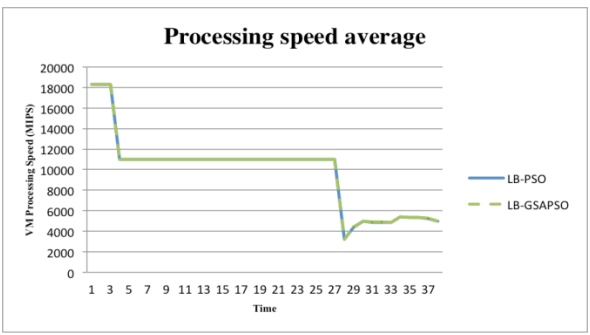

Fig. 2. Average VM processing speed over time.

## XII. CONCLUSION

In this paper, we have proposed a load balancing task scheduling algorithm for cloud computing environments based on the binary hybrid gravitational search and particle swarm optimization strategy. It balances the load of application requests submitted from cloud users over virtual machines in the cloud. The proposed algorithm enhances the overall VM utilization of the cloud system. We have compared our proposed hybrid algorithm with the pure Bin-LB-PSO. Results show that as the load increases over time, the processing speed of submitted applications decreases, which proves that the proposed Bin-LB-PSOGSA is more efficient in keeping the load balanced over time.

In the future, we plan to extend this kind of load balancing for workloads with dependent tasks. Also, we plan to improve this algorithm by considering other QoS factors, as well.